\documentclass[aps,prd,eqsecnum,showpacs,superscriptaddress,twocolumn,
floatfix,preprintnumbers,altaffilletter]{revtex4}
\usepackage{graphicx,url,amssymb,amsmath,longtable,rotating,color,units,wasysym,subfigure,wrapfig,epsfig}

\begin{document}

\title{Measuring the non-Gaussian stochastic gravitational-wave background: a method for realistic interferometer data}

\author{Eric~Thrane}
\email{ethrane@ligo.caltech.edu}
\affiliation{LIGO Laboratory, California Institute of Technology, MS 100-36, 
Pasadena, CA, 91125, USA}

\begin{abstract}
  A stochastic gravitational-wave background (SGWB) can arise from the superposition of many independent events.
  If the rate of events per unit time is sufficiently high, the resulting background is Gaussian, which is to say that it is characterized only by a gravitational-wave strain power spectrum.
  Alternatively, if the event rate is low, we expect a non-Gaussian background, characterized by intermittent sub-threshold bursts.
  Many experimentally accessible models of the SGWB, such as the SGWB arising from compact binary coalescences, are expected to be of this non-Gaussian variety.
  Primordial backgrounds from the early universe, on the other hand, are more likely to be Gaussian.
  Measuring the Gaussianity of the SGWB can therefore provide additional information about its origin.
  In this paper we introduce a novel maximum likelihood estimator that can be used to estimate the non-Gaussian component of an SGWB signature measured in a network of interferometers.
  This method can be robustly applied to spatially separated interferometers with colored, non-Gaussian noise.
  Furthermore, it can be cast as a generalization of the widely used stochastic radiometer algorithm.
\end{abstract}

\maketitle

\section{Introduction}
A stochastic gravitational-wave background (SGWB) is expected to arise from the superposition of many systems, which are individually too weak to detect, but which combine to produce a GW signature characterized by its ensemble statistical properties.
In astrophysical models~\cite{regimbau}, an SGWB can arise from objects such as compact binary coalescences~\cite{phinney,kosenko,zhu,StochCBC,RegPacCBC}, neutron stars (including highly magnetized neutron stars)~\cite{cutler,RegMan,rosado_ns,RegPac,howell2010,marassi}, young or spun-up neutron stars~\cite{RegPac,owen,barmodes1,barmodes2,barmodes3}, core collapse supernovae~\cite{firststars,howell2004,buonanno2005,marassi2009}, and white dwarf binaries~\cite{phinney_whitedwarfs}.
Cosmological/primordial sources, meanwhile, can arise from inflationary physics~\cite{grishchuk,starob,eastherlim,peloso,sorbo}, cosmic strings~\cite{caldwellallen,DV1,DV2,cosmstrpaper,olmez1,olmez2}, and pre-Big-Bang models~\cite{PBB1,PBBpaper}.

The initial LIGO and Virgo experiments have yielded a number of constraints on the SGWB~\cite{stoch-S5,s5vsr1,sph_results} including limits on the energy density of GWs, which surpass indirect bounds from Big Bang nucleosynthesis and measurements of the cosmic microwave background.
A worldwide network of second-generation GW interferometers are expected to begin taking data in 2015~\cite{aLIGO2,aVirgo,GEO2,CLIO,LCGT}, and recent work~\cite{StochCBC} indicates that realistic astrophysical models can be probed with second-generation advanced interferometers---most notably, the SGWB arising from binary neutron star and binary black hole coalescences.

In the event of a detection, it may not be immediately clear {\em which} systems give rise to the observed SGWB.
The strain power spectrum provides one tool for disentangling different possible sources~\cite{paramest,seto_params}.
Also, measurements of the SGWB can be compared with measurements of GW transients in order to indirectly infer information about the SGWB~\cite{paramest}.
Finally, sky maps of GW power and tests of isotropy provide yet another means of characterizing different models~\cite{sph_methods,sph_results,olmez2}.

\begin{figure}[hbtp!]
  \psfig{file=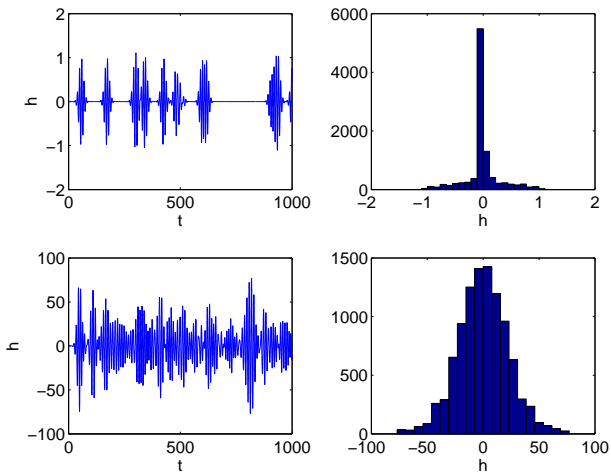, height=2.5in}
  \caption{
    An illustration of Gaussian and non-Gaussian signals.
    Top-left: a time series of sine-Gaussian bursts with a low duty cycle produces a non-Gaussian signal.
    Top-right: histogram of this non-Gaussian signal.
    Bottom-left: a time series of sine-Gaussian bursts with a high duty cycle produces an approximately Gaussian signal.
    Bottom-right: histogram of this Gaussian signal.
  }
  \label{fig:gaussian_example}
\end{figure}

In this paper we explore the {\em Gaussianity} of the SGWB.
A Gaussian SGWB is described only by its strain power spectrum, while a non-Gaussian SGWB (sometimes referred to as an SGWB in the ``popcorn'' or ``shot noise'' regime) consists of a series of discrete sub-threshold bursts~(see Figure~\ref{fig:gaussian_example}).
Non-Gaussian signals can be described with a probability distribution of burst waveforms and a duty cycle $\xi$, which we define as the fraction of data segments during which a GW source somewhere in the universe emits GWs in some analysis band.
In the analysis that follows, it will be useful to divide the full GW observing band into smaller analysis bands.

In this work we assume the duty cycle is less than unity, i.e., it is rare for two or more events to simultaneously emit in the same band.
For many astrophysical models, this is a good approximation.
In the case of the SGWB from binary neutron star coalescence, for example, we expect that $\xi\approx0.5\%$ for a $\unit[4]{Hz}$-wide bin centered at $\unit[100]{Hz}$.
We refer to both Gaussian and non-Gaussian signals as ``stochastic'' since both can be described in terms of the ensemble behavior of many individually undetectable bursts.
For a more nuanced discussion of relevant terminology, see Section~\ref{terminology} in the appendix.

Different models of the SGWB predict different levels of Gaussianity~(see, e.g.,~\cite{tania_strings}).
Early-universe scenarios, as a rule of thumb, produce nearly Gaussian signatures whereas astrophysical scenarios tend to be more non-Gaussian~\cite{rosado}.
Moreover, a single model can produce a range of different duty cycles and burst amplitudes depending on its parameters.
Measurements of SGWB Gaussianity can provide an important probe to distinguish between models and also to estimate cosmological parameters such as the GW burst rate.

We introduce a maximum likelihood statistic to estimate the duty cycle and other parameters associated with non-Gaussian SGWB signatures in GW interferometers.
In~\ref{radiometer}, we show that this statistic can be cast as a generalization of the stochastic radiometer~\cite{radio_method,ballmer}.
Previous work has yielded a nearly optimal technique (in the statistical sense) for the special case of colocated, co-aligned interferometers characterized by stationary, Gaussian white noise~\cite{drasco}.
More recently, Seto has proposed the use of higher order moments for measuring the Gaussianity of the SGWB~\cite{seto,setoBBO}.

Real GW interferometer data, however, is far from idealized noise.
It is colored, non-Gaussian, non-stationary, and colocated interferometers suffer from correlated noise.
These factors have make it challenging to implement the method from~\cite{drasco} in practice.
The maximum likelihood method proposed here is applicable to realistic GW interferometer noise.

In Section~\ref{formalism}, we review the canonical framework for analyzing the Gaussian SGWB and introduce a new formalism for handling non-Gaussian signals; in Section~\ref{simulation}, we demonstrate the non-Gaussian formalism with a Monte Carlo simulation; in Section~\ref{sensitivity} we discuss the sensitivity of SGWB searches using the non-Gaussian formalism.
Finally, in Section~\ref{conclusions}, we summarize our findings and describe the next steps necessary to implement the non-Gaussian formalism with actual data and for specific models.

\section{Formalism}\label{formalism}
\subsection{Gaussian searches}
In order to motivate our non-Gaussian statistic, we begin by describing how traditional (Gaussian) SGWB analyses are carried out.
The basic idea behind a Gaussian SGWB measurement is to cross-correlate strain time series from a pair of detectors $I$ and $J$ to create a cross correlation statistic, which is an estimator for the energy density spectrum of GWs (see, e.g.,~\cite{allen-romano,stoch-S5}):
\begin{equation}
  \Omega_\text{GW}(f) = \frac{f}{\rho_c} \frac{d\rho_\text{GW}}{df} .
\end{equation}
Here $f$ is the GW frequency, $d\rho_\text{GW}$ is the energy density of GWs in a frequency band $(f,f+df)$, and $\rho_c$ is the critical energy density of the universe.
By summing data from many time segments and frequency bins, it is possible to observe signals orders of magnitude smaller than the instantaneous noise curve.
For additional details, the interested reader is referred to~\cite{allen-romano}.

The SGWB is assumed to be isotropic, stationary, and Gaussian~\cite{allen-romano,christensen_prd}.
The measured strain time series in detector $I$ can be written as
\begin{equation}\label{eq:s}
  s_I(t) = h_I(t) + n_I(t) .
\end{equation}
Here $h_I(t)$ and $n_I(t)$ are respectively the astrophysical strain signal and the strain noise.
For spatially separated interferometers, $n_I(t)$ and $n_J(t)$ are expected to be uncorrelated.
Signal and noise are, of course, expected to be uncorrelated.
The signals in $I$ and $J$, however, are expected to be highly correlated:
\begin{equation}\label{eq:twopoint}
  \langle \tilde{h}^\star_I(t;f) \tilde{h}_J(t;f) \rangle \neq 0 .
\end{equation}
Here, tildes denote discrete Fourier transforms, and $(t;f)$ are spectrogram indices for time and frequency bins respectively.

\begin{figure}[hbtp!]
  \psfig{file=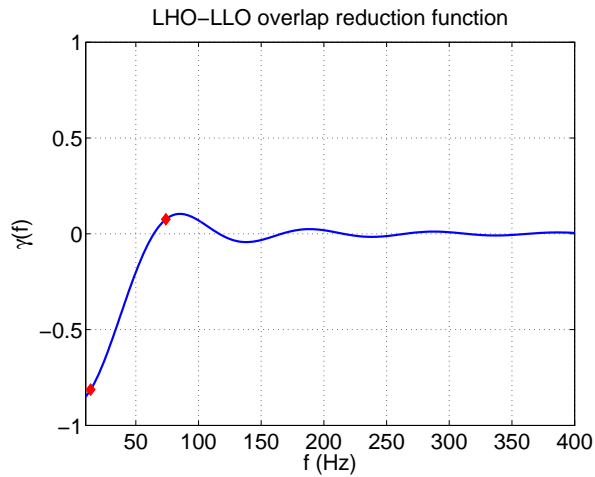, height=2.5in}
  \caption{
    The overlap reduction function $\gamma(f)$ for the LIGO Hanford - LIGO Livingston detector pair.
    The red diamonds mark frequency bins of $f=\unit[14]{Hz}$ and $\unit[74]{Hz}$, which are singled out below to illustrate how the distribution of non-Gaussian signal varies with frequency.
  }
  \label{fig:orf}
\end{figure}

From Eq.~\ref{eq:s}-\ref{eq:twopoint}, we can construct an estimator from the cross-power spectrum of the $I$ and $J$ strain channels~\cite{allen-romano}:
\begin{equation}\label{eq:Y}
  \hat{Y}(t;f) = 
  \frac{2}{T}
  Q(f) \text{Re} \left[ \tilde{s}^\star_I(t;f) \tilde{s}_J(t;f) \right] .
\end{equation}
$T$ is the segment duration and $Q(f)$ is a filter function, 
\begin{equation}
  Q(f) \equiv \frac{f^3}{\gamma(f) \Omega_M(f)}
\end{equation}
which emphasizes certain frequency bins depending on the spectral shape $\Omega_M(f)$ of a particular model, and a geometric factor called the overlap reduction function $\gamma(f)$ (see~Figure~\ref{fig:orf}), which takes into account the fact that uncorrelated GW signals from different parts of the sky interfere to reduce the observed correlation between spatially separated interferometers.

An estimator for the variance of $\hat{Y}(t;f)$ is given by
\begin{equation}
  \hat\sigma^2(t;f) = \frac{|Q(f)|^2}{2T\delta{f}} P_I(t;f) P_J(t;f) ,
\end{equation}
where $\delta{f}$ is the frequency resolution and $P_I(t;f)$ is the auto-power in detector $I$ in time-frequency bins $(t;f)$~\footnote{Typically, auto-power is calculated using adjacent time bins in order to avoid a bias that results from using the same data to calculate both $\hat{Y}(t;f)$ and $\hat\sigma(t;f)$}.
Armed with $\hat{Y}(t;f)$ and $\hat\sigma(t;f)$, the optimal estimator for $\int df \, \Omega_\text{GW}(f)$ at time $t$ is given by a weighted average:
\begin{equation}
  \begin{split}
    \hat{Y}(t) = \sum_f \hat{Y}(t;f) \, \hat\sigma^{-2}(t;f) / 
    \sum_f \sigma^{-2}(t;f)\\
    \hat\sigma^{-2}(t) = \sum_f \hat\sigma^{-2}(t;f) .
  \end{split}
\end{equation}
The optimal estimator for the entire data-taking period is a weighted average over time:
\begin{equation}\label{eq:Ytot}
  \begin{split}
    \hat{Y}_\text{tot} = \frac{\sum_t \hat{Y}(t) \, \hat\sigma^{-2}(t)}
        {\sum_t \hat\sigma^{-2}(t)} \\
        \hat\sigma_\text{tot}^{-2} = \sum_t \hat\sigma^{-2}(t) \\
        \text{SNR}_\text{tot} = \hat{Y}_\text{tot}/\hat\sigma_\text{tot} .
  \end{split}
\end{equation}
$\text{SNR}_\text{tot}$ is expected to be well-approximated by a normal distribution by the central limit theorem, and indeed, this is born out empirically~\cite{stoch-S5,sph_results}.

\subsection{A non-Gaussian statistic}
A number of applications have emerged that make use of $\hat{Y}(t;f)$ and $\hat\sigma(t;f)$---the intermediate data products that go into the calculation of $\hat{Y}_\text{tot}$ and $\hat\sigma_\text{tot}$~\cite{stamp,stamp_glitch,stamp_pem}.
Recent work utilizes spectrograms of cross- and auto- power to search for long-lived GW transients and to identify sources of environmental noise contaminating strain data channels.
Building on this work, we cast our non-Gaussian statistic in terms of these intermediate data.

To begin, we define a complex estimator
\begin{equation}
  \hat{Y}'(t;f) = Q(f) \tilde{s}^\star_I(t;f) \tilde{s}_J(t;f) ,
\end{equation}
which is a simple generalization of Eq.~\ref{eq:Y}.
Unlike a Gaussian background, which is isotropic at every instant in time, non-Gaussian bursts are associated with individual sky locations, (even if, on average, they are drawn from an isotropic distribution).
This information is encoded in the phase of $\hat{Y}'(t;f)$, and so it is necessary to work with both real and imaginary components.

We take as our starting point $\hat{Y}'(t;f)$ and $\hat\sigma(t;f)$ and their ratio
\begin{equation}\label{eq:rho}
  \rho(t;f) = \hat{Y}'(t;f) / \hat\sigma(t;f) .
\end{equation}
The quantity $\rho(t;f)$ has useful properties for our purposes.
First, it is well-studied and already in use in stochastic analyses.
Second, its statistical behavior can be probed robustly through time slides in which one strain time series is offset by an amount greater than the GW travel time between detectors in order to obtain many independent realizations of noise.

In order to derive our non-Gaussianity statistic, we endeavor to answer a simple question: how does the signal distribution of $\rho(t;f)$, denoted $S$, differ from the background distribution of of $\rho(t;f)$, denoted $B$.
We take each value of $(t;f)$---pixels in spectrograms of $\rho(t;f)$---to be a separate measurement.

For our purposes, {\em any measurement in which there is a non-Gaussian burst signal in $(t;f)$ is drawn from $S$}.
All other measurements are considered to be background.
This definition of signal and background is useful, but it can be counterintuitive.
The signal distribution, as we have defined it, is determined not only by properties of the non-Gaussian SGWB; it is also determined in part by properties of the detector noise.
Defining signal and background like this will be useful to derive an estimator for duty cycle.

Having pointed out these subtleties, we can now write the signal distribution as
\begin{equation}
  S(\rho(t;f) | \vec\theta_\text{signal}, \vec\theta_\text{noise}) ,
\end{equation}
where $\vec\theta_\text{signal}$ is a vector of parameters describing the SGWB and $\vec\theta_\text{noise}$ describes the detector noise.
In the same vein, we can write the background distribution as
\begin{equation}
    B(\rho(t;f) | \vec\theta_\text{noise}) .
\end{equation}
As we proceed we shall refer to simply $S(\vec\Theta)$ and $B(\vec\Theta)$, using capital $\vec\Theta$ as shorthand for the appropriate vector of parameters.
Further, we drop $(t;f)$ arguments in favor of $i$, which denotes separate measurements.
We return later to asymmetries in the $t$ and $f$ variables.

\begin{figure*}[hbtp!]
  \begin{tabular}{cc}
    \psfig{file=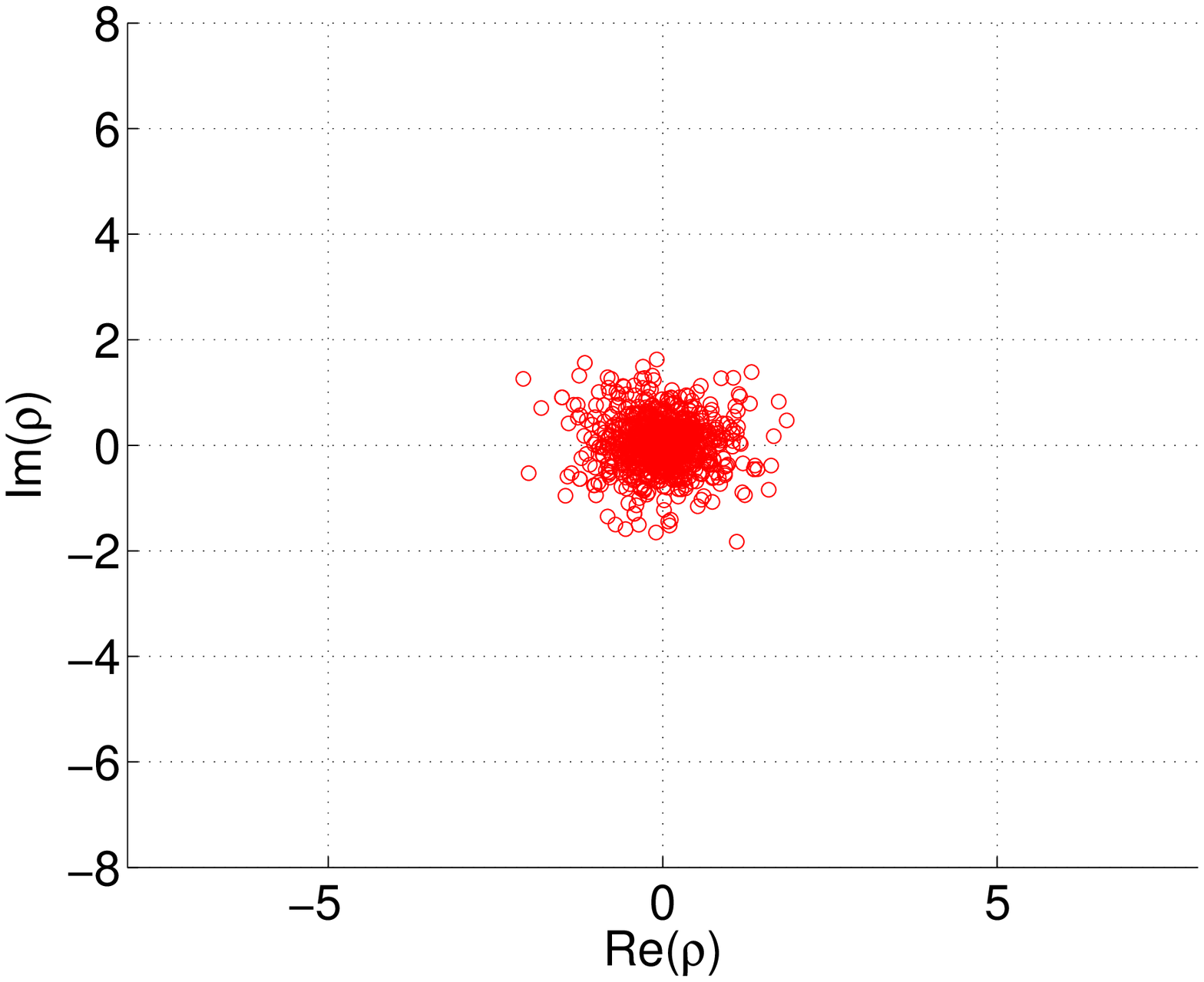, height=2.5in} &
    \psfig{file=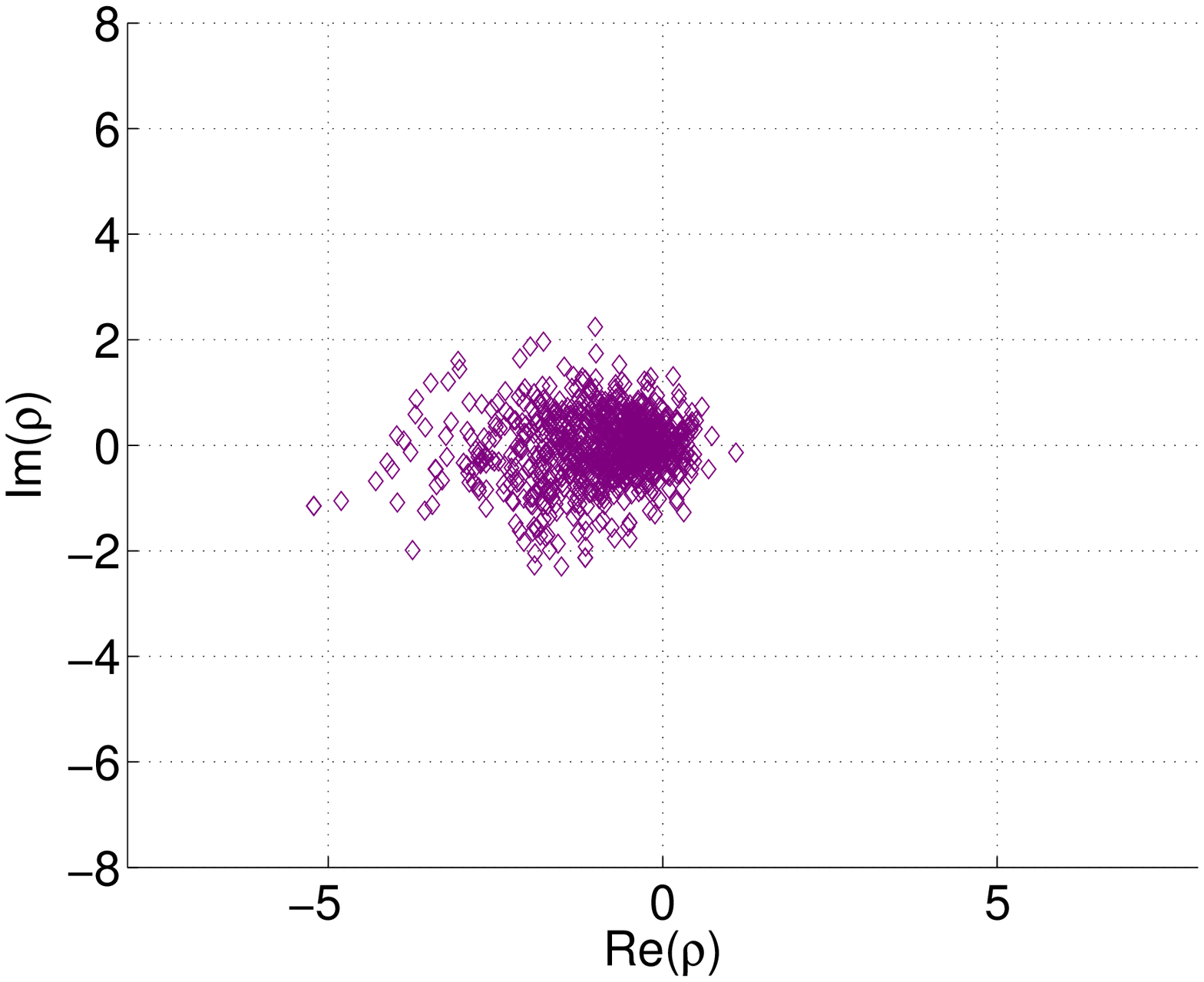, height=2.5in} \\
    \psfig{file=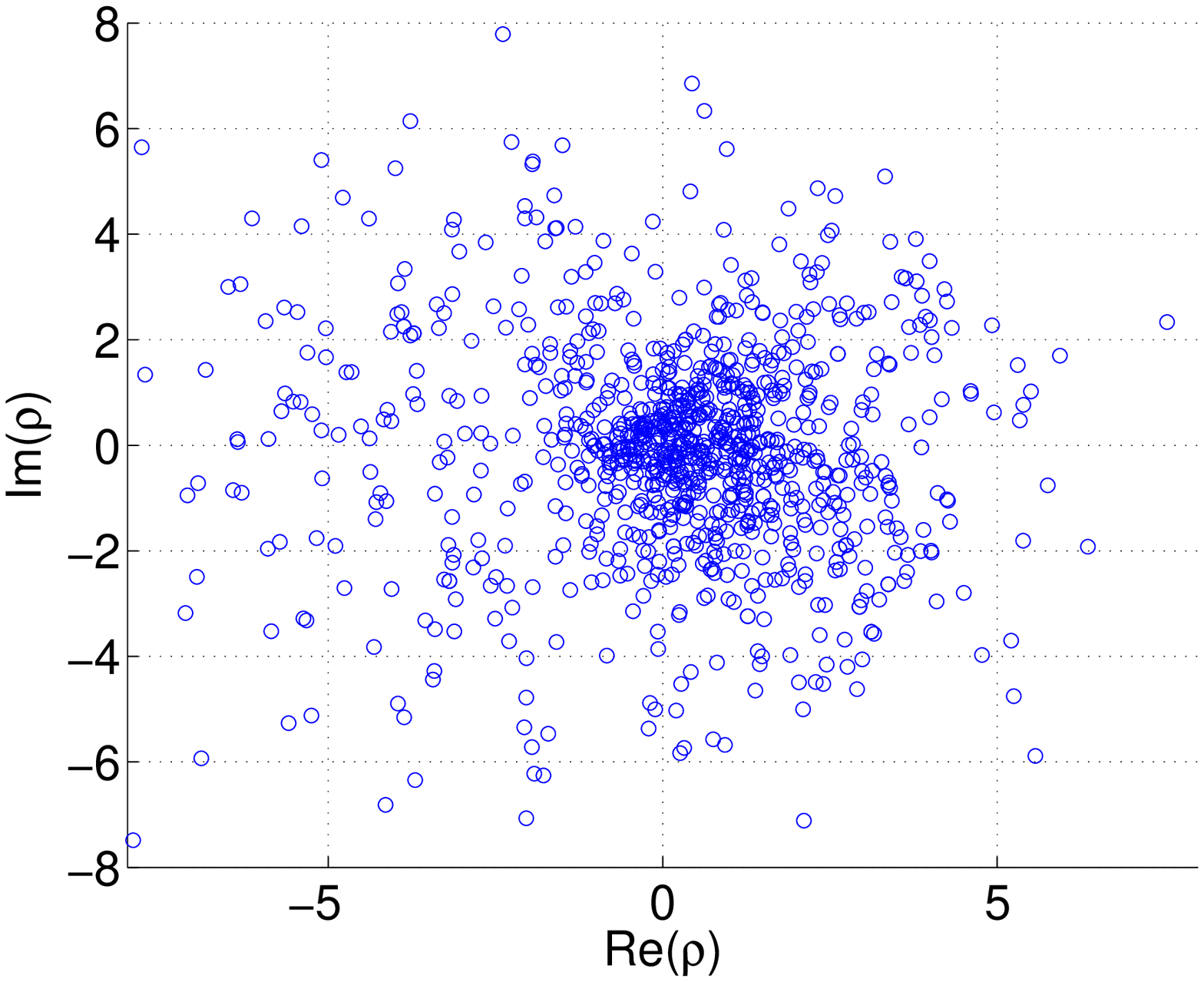, height=2.5in} &
    \psfig{file=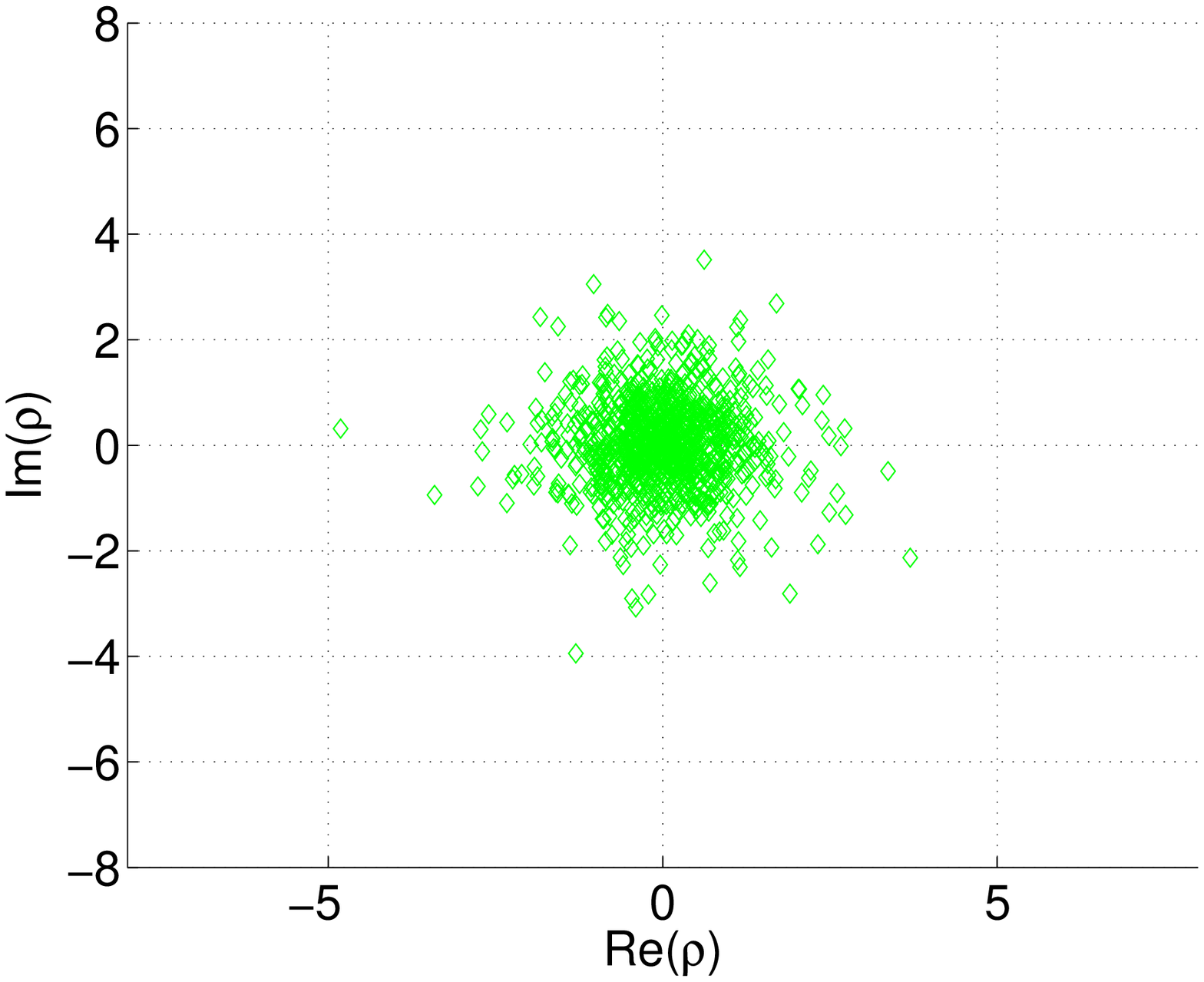, height=2.5in}
  \end{tabular}
  \caption{
    Scatter plots of $\rho$ for different signal and background models.
    Top-left: Gaussian noise with no signal (arbitrary frequency band).
    Top-right: Gaussian noise in the presence of a non-Gaussian signal; ($f=\unit[14]{Hz}$).
    Bottom-left: Gaussian noise in the presence of a non-Gaussian signal ($f=\unit[74]{Hz}$).
    Bottom-right: Gaussian noise with non-Gaussian glitches identical to the non-Gaussian signals, but only in one detector during each measurement ($f=\unit[74]{Hz}$).
    In each plot, the frequency bin width is $\unit[4]{Hz}$.
  }
  \label{fig:distributions}
\end{figure*}

Examples of $S(\rho_i|\vec\Theta)$ and $B(\rho_i|\vec\Theta)$ obtained using Monte Carlo are shown in Figure~\ref{fig:distributions}.
A detailed description of the Monte Carlo simulation is provided in the subsequent section.
For now, we simply describe the panels of Figure~\ref{fig:distributions} in general terms and point out some of the interesting features.

The top-left panel shows a scatter plot of $\rho$ for Gaussian noise.
(In this case, we consider a frequency bin centered on $f=\unit[74]{Hz}$, but the distribution looks the same at all frequencies.)
The distribution has a mean of zero and and is narrow compared to both non-Gaussian signal and non-Gaussian noise.
The top-right panel, meanwhile, shows the case of a non-Gaussian signal in the presence of Gaussian noise in a frequency bin centered on $f=\unit[14]{Hz}$.
At $\unit[14]{Hz}$, the GW wavelength $\lambda\approx\unit[2\times10^4]{km}$ is large compared to the separation of the interferometers, $\approx\unit[3000]{km}$ for the case of the LIGO Hanford Observatory (LHO) and LIGO Livingston Observatory (LLO) as used in this simulation.
Thus, $\rho$ is distributed approximately as it would be for a colocated pair of interferometers.
The presence of a signal is evidenced by the shift of the distribution away from zero.
The fact that the shift is negative is due to relative orientations of the LHO and LLO detectors, which is encoded in the sign of the overlap reduction function (see Figure~\ref{fig:orf}).

The lower-left panel shows the distribution of $\rho$ for non-Gaussian signal in the presence of Gaussian noise in a frequency bin centered on $f=\unit[74]{Hz}$.
At $\unit[74]{Hz}$, the interferometers no longer behave as though they are colocated.
The presence of a signal shifts the mean away from zero, but it also changes the width and shape of the distribution.
Since $\unit[74]{Hz}$ occurs in between the first and second zeros of the overlap reduction function, the mean is positive (see Figure~\ref{fig:orf}).
For illustrative purposes, the non-Gaussian bursts used to make this plot are made loud compared to the detector noise.

The lower-right panel shows non-Gaussian detector noise (also called ``glitchy'' noise) simulated by taking the same bursts used to simulate a non-Gaussian signal (lower-left), but generating a signal in only one detector at a time; i.e., we assume the noise at each site is uncorrelated.
The glitchiness of the signal widens the distribution compared to Gaussian noise, but not as much as coincident non-Gaussian signals, which suggests that we can differentiate population of glitches from non-Gaussian signals.
Additionally, the mean of the non-Gaussian noise distribution is still zero, like Gaussian noise.
By comparing these distributions of different signal and noise distributions, we can build a statistical framework to differentiate them.
This is the goal of the remainder of this section.

In practice, $S(\rho_i|\vec\Theta)$ and $B(\rho_i|\vec\Theta)$ can be calculated from Monte Carlo or from pseudo experiments performed with time-shifted data.
The former method is computationally cheaper, but yields a less accurate description of the noise.
Even an approximate Monte Carlo description, however, can be a useful starting point.
$S(\rho_i|\vec\Theta)$ and $B(\rho_i|\vec\Theta)$ are used to weight data as more or less like signal.
To the extent that they differ from the true distributions, the likelihood statistic will be less effective distinguishing between signal and background.
However, if the likelihood statistic is ultimately tested empirically on time-shifted data, then we can avoid bias in detection or parameter estimation---even if we construct the estimator with an only approximate Monte Carlo model.

Armed with distributions of of signal and background, the likelihood of observing $\rho_i$ can be written as
\begin{equation}\label{eq:P}
  P_i(\xi|\rho_i,\vec\Theta) = 
  \xi S(\rho_i|\vec\Theta) + (1-\xi) B(\rho_i|\vec\Theta) .
\end{equation}
Here $\xi$ is the probability that the measurement is drawn from the signal distribution.
We can interpret $\xi$ as the duty cycle for our non-Gaussian signal model.
This formulation is similar to a maximum likelihood approach used in neutrino and gamma-ray astronomy, see, e.g.,~\cite{skptsrc,grb080319B,braun}.

Given $N$ measurements, we can construct a likelihood function 
\begin{equation}\label{eq:likelihood}
  {\cal L}(\xi|\{\rho_i\},\vec\Theta)=\prod_i^N P_i(\rho_i|\rho_i,\vec\Theta) .
\end{equation}
Here, we have implicitly assumed that each measurement is statistically independent.
This assumption is expected to be approximately valid for reasonably stationary noise and for signals that do not repeatedly wander in and out of the same frequency bin.
Subtle effects from overlapping data segments may require a more careful treatment.
An estimator for $\xi$ (denoted $\hat\xi$) is given simply by maximizing ${\cal L}(\xi)$.
Confidence intervals for $\xi$ are calculated straightforwardly from ${\cal L}(\xi|\{\rho_i\},\vec\Theta)$.
To illustrate, we perform a simple Monte Carlo calculation, described in detail in Section~\ref{simulation} and summarized in Figure~\ref{fig:example_intervals}.

Finally, we consider the question of how best to detect non-Gaussianity in an SGWB signal.
We can frame this question more precisely as: what is the appropriate metric for determining whether $\xi$ is non-zero.
A convenient measure of non-Gaussian signal strength is the ratio of the likelihood evaluated at its maximum to the likelihood evaluated at $\xi=0$ (see~\cite{skptsrc,grb080319B,braun}):
\begin{equation}
  \Lambda = 2 \log\left[{\cal L}(\hat\xi) / {\cal L}(0)\right] .
\end{equation}
If $S(\rho_i|\vec\Theta)$ and $B(\rho_i|\vec\Theta)$ are accurate descriptions of the signal and background, then the probabilistic interpretation of $\Lambda$ as (twice the log of) a likelihood ratio is straightforward.
However, even if $S(\rho_i|\vec\Theta)$ and $B(\rho_i|\vec\Theta)$ are only approximately known, we can calculate $\Lambda$ for many realizations of time-shifted data.
In this way, we can obtain a robust and empirical means of converting $\Lambda$ to a false alarm probability---even if our noise and signal models are only roughly approximate.

\subsection{Relationship to radiometer statistic}\label{radiometer}
The stochastic radiometer statistic~\cite{radio_method,ballmer} applies the Gaussian isotropic formalism of~\cite{allen-romano} to the case of a Gaussian point source.
In this subsection we show how the radiometer statistic can be cast as a special case of our non-Gaussian statistic.
We begin by defining the signal model.
We consider a point source associated with a fixed sky location $\hat{n}$.
We assume that the source is characterized by a stationary GW energy density spectrum $\Omega(f)=\overline\Omega$.
(For the sake of simplicity, we assume that it is constant with respect to $f$.)

For a single point source, there is a known phase relationship between $I$ and $J$ and so Eq.~\ref{eq:Y} becomes~\cite{radio_method}
\begin{equation}
  \hat{Y}(t;f) = Q(f) \text{Re}\left[
    e^{-2\pi i f \tau(\hat{n},t)}
    \tilde{s}_I^\star(t;f) \tilde{s}_J(t;f)
    \right] .
\end{equation}
Here $\tau(\hat{n},t)$ is the delay time between the interferometers.
The signal distribution is given approximately by~\cite{stamp}
\begin{equation}
  S(\rho_i | \overline\Omega) \propto 
  \exp\left(-\frac{1}{2}(\rho_i - \overline\Omega/\sigma_i)^2\right) .
\end{equation}
Since the source is, by assumption, always emitting GWs, we can set $\xi=1$, which implies that there is GW signal present in every data segment and that we can ignore $B(\rho_i|\vec\Theta)$ entirely.
It follows from Eqs.~\ref{eq:P}-\ref{eq:likelihood} that the likelihood can be written as
\begin{equation}
  \begin{split}
  {\cal L}(\xi=1|\{\rho_i\}, \overline\Omega) \propto 
  \exp\left( -\frac{1}{2}\sum_i^N (\rho_i-\overline\Omega/\sigma_i)^2 \right)\\
  \propto
  \exp \left( -(\overline\Omega-\hat{Y}_\text{tot})^2 / 2\sigma_\text{tot}^2 \right) , 
  \end{split}
\end{equation}
where $\hat{Y}_\text{tot}$ is simply the optimal estimator from Eq.~\ref{eq:Ytot}.

Thus, the radiometer statistic is a special case of the non-Gaussian statistic in which the signal model fixes the sky location of the source and the duty cycle is set to $\xi=1$.
The likelihood function can then be used to estimate the energy density spectrum $\overline\Omega$.
In principle, a similar analogy is possible between the isotropic statistic~\cite{allen-romano} and the non-Gaussian statistic.
However, the fact that both the radiometer and the non-Gaussian statistic assume GW point sources makes the analogy shown here more straightforward.

\subsection{Comparison to other methods}
One of the first papers to address the topic of detecting a non-Gaussian SGWB is~\cite{drasco}.
Our method differs in several important ways.
In this work we relax the assumptions from ~\cite{drasco} that the noise is Gaussian and white.
Instead of relying on two colocated detectors as in~\cite{drasco}, we assume two spatially separate detectors.
Unlike~\cite{drasco}, our likelihood statistic is framed in terms of cross-power, with auto-power terms used only for normalization.

Since the statistic in~\cite{drasco} is nearly optimal, it is very likely to provide a more sensitive measurement within its domain of utility, compared to the method described here.
However, our statistic (built from cross-power) can be extended straightforwardly to interferometers with colored, non-Gaussian noise, and they need not be colocated.
Spatially separated interferometers, in turn, allow for the use of robust background estimation through time slides.

Statistics utilizing forth-order (kurtosis) strain moments have been proposed as probes for non-Gaussianity in the SGWB~\cite{seto,setoBBO}.
It is presently difficult to evaluate the relative merits of different non-Gaussianity techniques, since none has been utilized with real interferometer data.
Clearly, the next step is to carry out analyses with real data.

\section{Simulation and Results}\label{simulation}
In order to demonstrate our likelihood formalism, we perform a Monte Carlo simulation.
We generate three types of data: Gaussian noise, Gaussian noise + non-Gaussian GW bursts (signal), and non-Gaussian noise.
The Gaussian noise is colored according to the design sensitivity of Advanced LIGO~\cite{aLIGO2} and we assume a network consisting of $\unit[4]{km}$ detectors at LHO and LLO.
We employ a toy signal model consisting of randomly arriving $\unit[200]{ms}$ white-noise bursts with a strain amplitude density of $\approx\unit[3\times10^{-24}]{Hz^{-1/2}}$.
These bursts are marginal compared to the noise---the average $|\hat\rho|$ is $0.43$ for bursts plus noise and only slightly less ($0.41$) for noise alone---and can therefore be characterized as sub-threshold.

We calculate spectrograms of $\rho(t;f)$ (Eq.~\ref{eq:rho}) using $\unit[4]{Hz} \times \unit[1]{s}$ pixels.
Since we are presently concerned only with demonstration, this choice of pixel size is arbitrary.
The issue of pixel size is revisited in the appendix.

\begin{figure}[hbtp!]
  \psfig{file=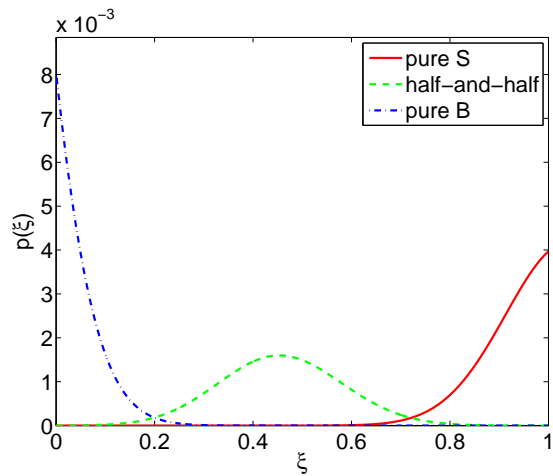, height=2.5in}
  \caption{
    Example posteriors for duty cycle $\xi$ using Monte Carlo data for pure background $\xi=0$ (dash-dot blue), pure signal $\xi=1$ (solid green), and an even mixture of the two $\xi=0.5$ (dashed green).
  }
  \label{fig:example_intervals}
\end{figure}

In order to construct the likelihood function used in Figure~\ref{fig:example_intervals}, we construct distributions of $S(\rho_i|\vec\Theta)$ and $B(\rho_i|\vec\Theta)$ using $10^7$ trials of Monte Carlo data.
We then use an independent dataset consisting of $500$ Monte Carlo measurements, which---following Eq.~\ref{eq:likelihood}---we compare to $S(\rho_i|\vec\Theta)$ and $B(\rho_i|\vec\Theta)$ in order to measure the duty cycle $\xi$~\footnote{Using this pixel resolution, one year's worth of data would correspond to a significantly larger dataset---$\approx3\times10^7$ measurements for every frequency bin---so this demonstration is not meant to be representative of a typical year-long stochastic search.}.

The results of our simulation are illustrated in Figure~\ref{fig:example_intervals}.
We test three datasets: one consisting of pure background ($\xi=0$, dash-dot blue), one consisting of pure signal ($\xi=1$, solid red), and a third consisting of a $50\%$ mixture of each ($\xi=0.5$, dashed green).
In all three cases we find that the observed posterior distributions are consistent with the true value of $\xi$.
This demonstrates, with a very simple toy model, how our formalism can be used to measure $\xi$ in the presence of colored noise in spatially separated interferometers.

\section{Sensitivity}\label{sensitivity}
A natural question is: if the non-Gaussian search presented here incorporates information about the non-Gaussian character of the popcorn signal we seek to measure, can it, in some cases, provide a more sensitive detection statistic than the Gaussian statistic used in previous stochastic searches~\cite{stoch-S5,s5vsr1,sph_results}?
A detailed analysis, beyond our present scope, is required to answer this question thoroughly.
However, there are several points worth noting.

First, in the limit of (highly idealized) stationary Gaussian noise, we expect the non-Gaussian statistic will outperform the Gaussian statistic.
To illustrate, we note that the green data in Figure~\ref{fig:example_intervals} have a Gaussian statistic signature of $\text{SNR}_\text{tot}<1$ (typical of pure noise) whereas the non-Gaussian statistic $\Lambda=15$ represents a strong detection.
We also expect, however, that as the data becomes glitchier, the advantage of the non-Gaussian approach will diminish, since both glitches and non-Gaussian bursts will have a tendency to perturb higher-order moments of the distribution of $\rho(t;f)$ (albeit in different ways).

Second, the Gaussian statistic is almost completely insensitive to stochastic signals in frequency bins corresponding to the zeros of the overlap reduction function $\gamma(f)$.
These zeros represent frequencies at which the detector pair are as likely to be out of phase as in phase, and so the integrated signal is zero.
Since the non-Gaussian technique presented here incorporates higher-order moments in the distribution of $\rho(t;f)$ (beyond the mean), it will have at least some sensitivity to the SGWB even when $\gamma(f)\approx0$.

Third, while there are potential advantages associated with the non-Gaussian statistic, it is worthwhile to mention several advantages possessed by the Gaussian statistic.
It is very well studied and has been shown to yield reliable results~\cite{stoch-S5,s5vsr1,sph_results,allen-romano,stoch-S1,stoch-S3,stoch-S4}, it is simple to understand and implement, and since it utilizes the sum of a great many numbers, it is very robust to non-stationary noise artifacts.
Thus, the Gaussian statistic is likely to provide an important benchmark and cross-check to results obtained with the non-Gaussian statistic.

\section{Conclusions}\label{conclusions}
Some of the most promising sources of the stochastic gravitational-wave background (such as compact binary coalescences) are likely to be non-Gaussian.
By measuring the non-Gaussianity of the stochastic background, we can learn more about its origin.
To this end, we have presented a maximum likelihood estimator that can be used to measure the non-Gaussianity of the stochastic gravitational-wave background utilizing realistic interferometer data.
Using Monte Carlo data, we illustrated how the calculation can be carried out, and demonstrated that we can estimate the duty cycle of the bursts that characterize a non-Gaussian signal.
We outlined the next steps, which must be undertaken in order to tune the analysis for specific astrophysical models such as the stochastic background arising from compact binary coalescences.
Future work will focus on carrying out this optimization.

\begin{acknowledgments}
  We thank Joseph Romano and Tania Regimbau for thoughtful comments on a draft of this paper.
  This work is by a member of the LIGO Laboratory, supported by funding from United States National Science Foundation.
  LIGO was constructed by the California Institute of Technology and Massachusetts Institute of Technology with funding from the National Science Foundation and operates under cooperative agreement PHY-0757058.
  This paper has been assigned LIGO document number LIGO-P1200141.
\end{acknowledgments}

\begin{appendix}
\section{Terminology}\label{terminology}
Rosado~\cite{rosado} has reviewed the SGWB literature and attempted to provide a comprehensive and standardized glossary of terminology.
Where possible, we try to follow the terminology of~\cite{rosado}, referring, for example, to the objects, which combine to create a SGWB as ``systems.''
However, while we are loath to add to the SGWB lexicon, some definitions and distinctions are necessary for our present purpose.

Rosado makes a distinction between (in-principle) {\em resolvable} and {\em unresolvable} SGWBs.
Unresolvable SGWB signals, according to~\cite{rosado}, are present when, on average, two or more systems simultaneously create strain signals in the same frequency bin.
This distinction is most useful in the context of a far-future detector with sufficient sensitivity to subtract out resolvable signals in order to measure an underlying primordial SGWB~\cite{cutler-harms}.
Near-future detectors will lack the sensitivity to separately measure the systems contributing to the ``resolvable'' SGWB.
It is therefore useful to define {\em sub-threshold bursts} as the components of an in-principle resolvable SGWB, which cannot be resolved in practice.
An SGWB consisting of sub-threshold bursts is always resolvable according to the definition in~\cite{rosado}.
Whether or not a resolvable SGWB consists of sub-threshold bursts will depend on the detector used to measure it.
The non-Gaussian SGWB considered in this paper consist of sub-threshold bursts.

\section{Details}
Here we point out details that will require more careful attention in order to implement this method for a specific SGWB model.
Our aim is not to provide a systematic treatment, but rather to highlight some of the finer points worthy of attention.

{\em Probability density functions.}---The distributions for $S(\rho_i|\vec\Theta)$ and $B(\rho_i|\vec\Theta)$ must be sampled with sufficient resolution to distinguish between signal and background.
Models with very low-level bursts may require very high resolutions, and so significant computational resources may be necessary to compute $S(\rho_i|\vec\Theta)$ and $B(\rho_i|\vec\Theta)$.

{\em Pixel size.}---Pixel size can be chosen to optimize the sensitivity of a search.
The pixel dimensions should be chosen so as to be comparable to the time and bandwidth of the non-Gaussian burst that is the target of the search.
Pixels that are very long/short in time are undesirable because, in the first case, the signal will be diluted with more noise than necessary, and in the second case, the signal will be spread thinly over many pixels.
An analogous argument can be made for the frequency bin width.
Numerical studies determine a suitable pixel size appropriate for a given model.

{\em Broadband analysis.}---The behavior of $S(\rho_i|\vec\Theta)$ and $B(\rho_i|\vec\Theta)$ varies significantly depending on the frequency band of interest.
For example, the mean of $S(\rho_i|\vec\Theta)$ can be positive, negative, or zero depending on the value of the overlap reduction function at the frequency in question (see Figure~\ref{fig:orf}).
(For bursts drawn from an isotropic distribution, $\langle \rho_i \rangle$ is always real since $\rho$'s drawn from some direction $\hat{n}$ have, on average, the opposite imaginary component of $\rho$'s drawn from the antipodal direction $-\hat{n}$.)
Therefore, it may be desirable to calculate $S(\rho_i|\vec\Theta)$ and $B(\rho_i|\vec\Theta)$ for many different bands.

Additional complications may arise from the fact that the signal may not spend the same duration emitting in every band.
Compact binary coalescences, for example, emit at a frequency that accelerates as a function of time.
In order to combine posterior distributions of $\xi$ from different bands, it may therefore be necessary to apply normalization factors to take into account the expected frequency evolution of the signal.

{\em Time-varying detector performance.}---For a variety of reasons, real GW detectors vary in performance on timescales ranging from minutes to months.
As examples, anthropogenic noise can cause elevated noise levels during the local rush hour, and noise performance can improve month to month following commissioning breaks.
In the example plots showed in Figure~\ref{fig:distributions}, we assume that the detector noise is stationary.

In order take into account the variability of the noise as a function of time, it may prove useful to add another variable to $\vec\Theta$ describing the variability of the noise.
Another option could be to simply use a subset of the highest quality data in which the strain sensitivity and glitchiness are relatively uniform.
\end{appendix}

\bibliography{popcorn}

\end{document}